\documentclass[prl,twocolumn,showpacs,preprintnumbers,amsmath,amssymb,nofootinbib]{revtex4}
\usepackage{subfigure}
\usepackage{graphicx}
\usepackage{bm}

\newcommand{\rd}{\textrm{d}}
\newcommand{\ee}{\begin{equation}}
  \newcommand{\eee}{\end{equation}}
\newcommand{\ea}{\begin{eqnarray}}
  \newcommand{\eea}{\end{eqnarray}}

\newcommand{\mpc}{\rm Mpc}

\newcommand{\od}{\Omega_d}
\newcommand{\osf}{\bar{\Omega}_{\rm sf}}
\newcommand{\ols}{\bar{\Omega}_{\rm ls}}
\newcommand{\ode}{\Omega_d^e}
\newcommand{\omh}{\Omega_m^0 h^2}
\newcommand{\obh}{\Omega_b^0 h^2}

\preprint{HD-THEP-05-17}
\begin{document}

\author{Michael Doran}
\affiliation{Institut f\"ur Theoretische Physik, Universit\"at Heidelberg,Philosophenweg 16, 69120 Heidelberg, Germany}
\author{Khamphee Karwan}
\affiliation{Institut f\"ur Theoretische Physik, Universit\"at Heidelberg,Philosophenweg 16, 69120 Heidelberg, Germany}
\author{Christof Wetterich}
\affiliation{Institut f\"ur Theoretische Physik, Universit\"at Heidelberg,Philosophenweg 16, 69120 Heidelberg, Germany}
\title{Observational constraints on the dark energy density evolution}
\begin{abstract}
We constrain the evolution of the dark energy density $\od$ from Cosmic Microwave Background, Large Scale Structure,
and Supernovae Ia measurements. While Supernovae Ia are most sensitive to the equation of state $w_0$ of dark
energy today, the Cosmic Microwave Background and Large Scale Structure data best constrains the dark energy
evolution at earlier times. For the parametrization used in our models, we find $w_0 < -0.8$ and the dark energy
fraction at very high redshift $\ode < 0.03$ at 95 per cent confidence level.
\end{abstract}
\pacs{98.80.-k}

\maketitle

Observations 
\cite{Riess:2004nr,Spergel:2003cb,Readhead:2004gy,Goldstein:2002gf,Rebolo:2004vp,Tegmark:2003ud,Hawkins:2002sg} indicate that a 
mysterious form of dark energy \cite{Wetterich:fm,Ratra:1987rm,Caldwell:1997ii,Caldwell:1999ew}
is driving an accelerated expansion of our Universe.
So far, the focus has been on the equation of state $w\equiv \bar p/\bar\rho$
of dark energy and in particular on its current value $w_0$. However,
if a dynamical dark energy or quintessence arises from
the time evolution of a scalar (cosmon) field one expects in 
general, that 
 the equation of state 
changes as the Universe expands. Various parameterizations of $w$ as a
function of the scale factor $a$ or redshift $z$ have been
investigated \cite{Corasaniti:2002vg,Linder:2002et,Upadhye:2004hh,Wang:2004py} . Yet, it is rather
the amount of dark energy $\od$ than the equation of state (which is
related to the derivative of $\od$) that influenced our Universe in the past.
 In this spirit, suitably averaged quantities like $\ols$
\cite{Doran:2000jt} and $\osf$ \cite{Doran:2001rw} have been used to
describe the effects of dark energy 
at the time of last scattering and 
during
 structure formation.

No parameterization of $w$ or $\od$ -- no matter how
complicated -- will perfectly describe the evolution of dark energy.
Yet, some essential key features for a viable model seem to be the
following: today, the amount of dark energy should be $\sim 70 \%$.
Going back in time, this value must have decreased considerably, as
current constraints yield a fraction of dark energy at the time of 
last scattering $\ols \lesssim 8 \%$ \cite{Caldwell:2003vp}. 
Supernovae measurements tell us that this decrease must have occurred swiftly, as
the slope of this decrease is reflected in $w_0 \lesssim -0.7$.

\begin{figure}
\includegraphics[width=0.45\textwidth,angle=0]{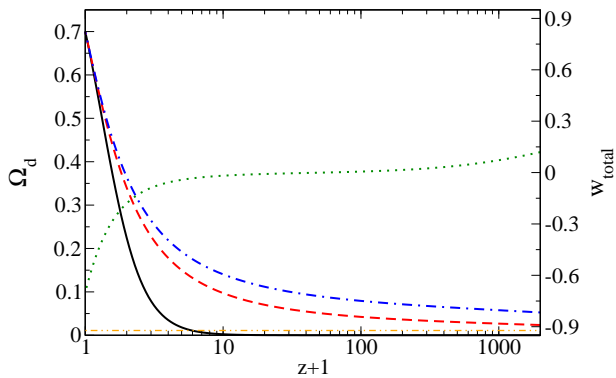}
\caption{Evolution of a typical dark energy model considered in this paper
with $\od^0 = 0.7$, $w_0 = -0.999$ and $\ode = 0.01$ depicted as dashed (red)
line and with $\ode = 0.03$ depicted as dashed-dotted (blue) line. These two models
correspond to values of $\ode$ that are allowed within $1\sigma$ and $2\sigma$ respectively. 
For comparison, the solid (black) line shows the evolution
for a cosmological constant with same $\od^0$. The straight dashed-double dotted 
(orange) line at $\od=0.01$
indicates the asymptotic limiting value of $\od$ for our model with $\ode=0.01$.
In contrast to the cosmological constant which contributes negligible at early
times, our models have $\od \geq 0.01\ (0.03)$ always and contribute a considerable effective
fraction $\osf \sim 0.06\ (0.1)$ during structure formation.
The dotted (green) line indicates the total equation of state of the Universe for our $\ode=0.01$ model. For
$w_{total} < -1/3$, the Universe is accelerating.}
\label{fig::illustration}
\end{figure}

\begin{figure}
\vspace{1.2ex}
\includegraphics[width=0.45\textwidth,angle=0]{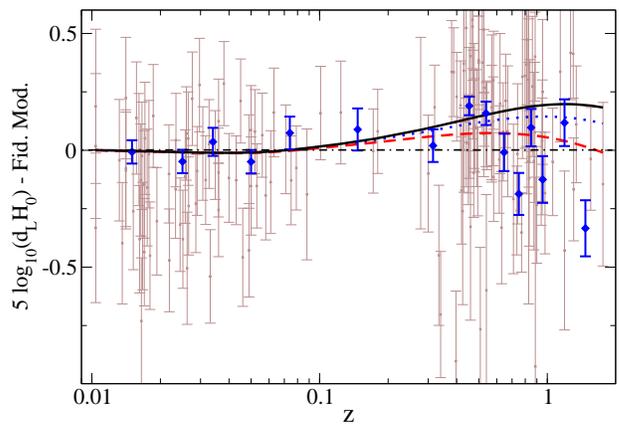}
\caption{Supernovae compilation of Riess et. al. \cite{Riess:2004nr} as data points
with thin (brown) error bars. In addition we plot the same data in binned 
fashion for illustration purposes \cite{leibundgut}. We plot
the logarithm of the luminosity distance minus a fiducial model for 
which $d_L H_0 = (1+z) \ln (1+z)$. The solid (black) line is for a cosmological
constant model, the dotted (blue) line is for $\ode = 10^{-4}$ and 
the dashed (red) line is for $\ode = 10^{-1}$. All models have $w_0 = -1$ in common.}
\label{fig::sne_overview}
\end{figure}

Usually, observations at low redshift, such as Sne Ia measurements are
combined with structure formation and CMB observations that are
probing earlier epochs. In this paper, we will take a different
point of view. Given the uncertainties in parameterizing $\od$ or
equivalently $w$, we look at high redshift and low redshift
constraints separately.
\begin{figure}
\includegraphics[width=0.21\textwidth,angle=-90]{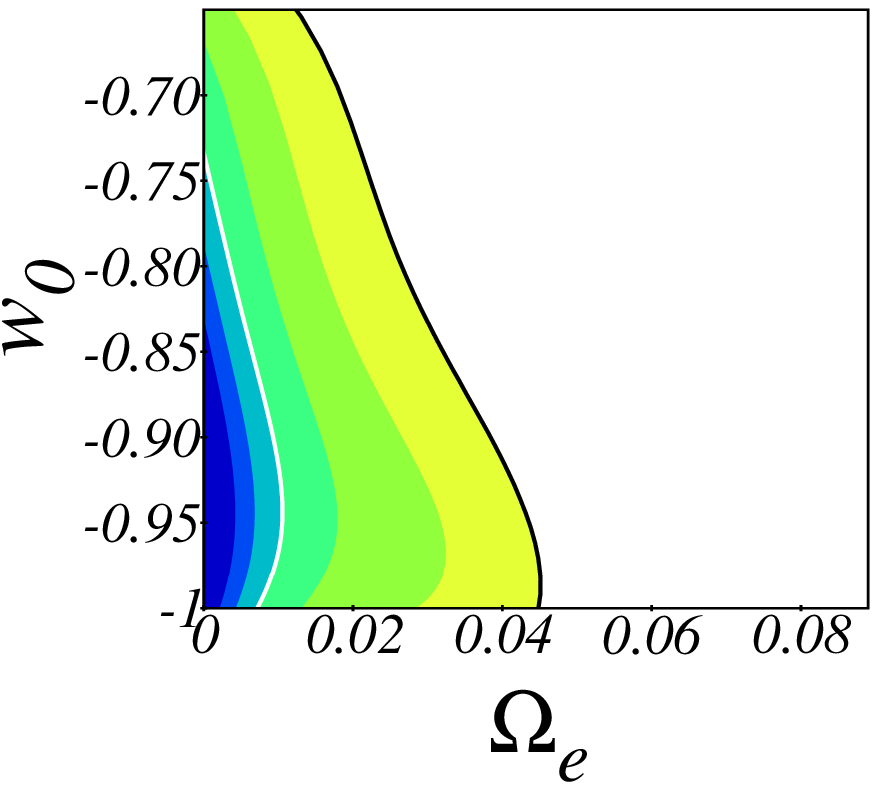}
\includegraphics[width=0.21\textwidth,angle=-90]{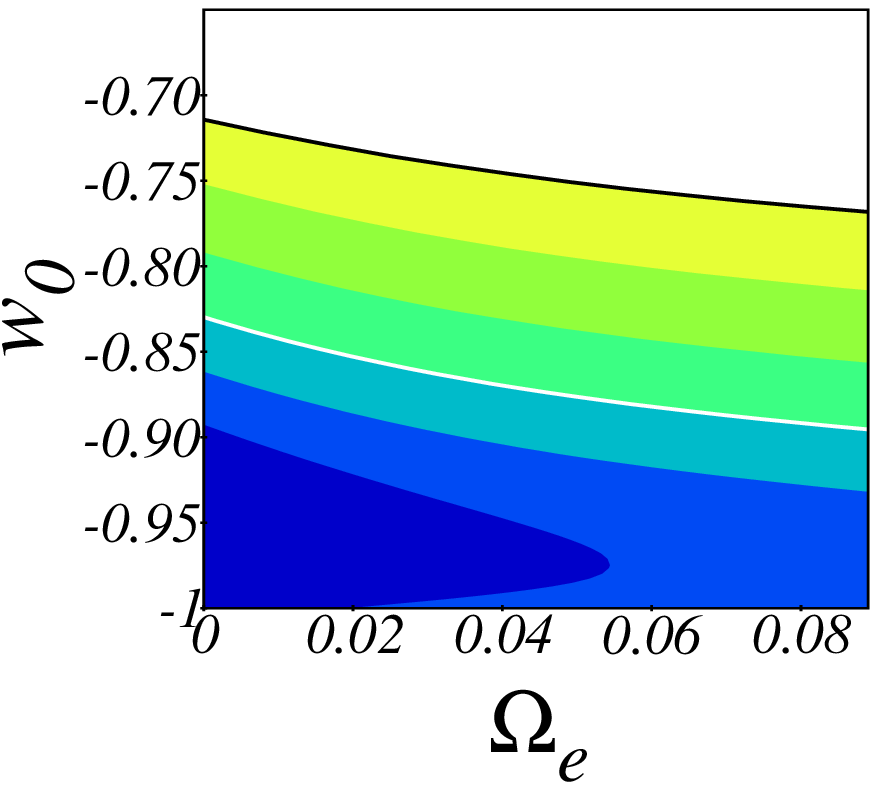}
\caption{Constraints on the model parameters $\ode$ and $w_0$. The left panel 
depicts the distribution from WMAP+CBI+VSA+SDSS+HST data, the right panel
that of Sne Ia alone. The regions of $68$\% ($95$\%)
are enclosed by a white (black) line. The two data sets give almost
orthogonal information: Sne Ia constrain $w_0$ but are less sensitive to $\ode$,
while CMB and LSS are more sensitive to $\ode$ (and hence $\ols$ and $\osf$)
than to the precise value of $w_0$.}
\label{fig::oe_w}
\end{figure} 
We use a particularly simple and direct parameterization of the dark
energy evolution \cite{Wetterich:2004pv}.  The parameters are the
amount of dark energy today $\od^0$ and the amount of dark energy at
early times $\ode$ to which it asymptotes for very large $z$.  In
terms of these, our parameterization is 
\ee \od(a) = \frac{e^R}{1 +
  e^R}, 
\eee 
where $R(a) \equiv \ln (\od(a) / [ 1 -\od(a)])$ obeys
\ee R(a) = R_0 - \frac{3 w_0 \ln a}{1 - b \ln a}. 
\eee
The constant $b$ is fully specified by the three parameters $w_0$,
$\od^0$ and $\ode$:
\ee b = -3 w_0 \left(\ln \frac{1 - \ode}{\ode} +
  \ln \frac{\od^0}{1-\od^0}\right)^{-1}.  
\eee
In terms of $w_0$ and $b$,  the equation of state is  $w(a) = {w_0}/{(1-b \ln a)^2}$.
For a comparison of our model to a cosmological constant, see figures \ref{fig::illustration} and \ref{fig::sne_overview}.
The advantage of a parameterization in terms of only
three parameters is a minimal setting that accounts for information
beyond a Taylor expansion around $z=0$ (i.e. beyond the parameters 
$\od^0$ and $w_0$). Nevertheless, it embodies the most crucial potential
new feature of a dynamical dark energy, namely the possibility of early dark 
energy. For supernovae, it seems more suitable than
a continuation of the Taylor 
expansion which at the next step involves the derivative
of the equation of state $w_0^\prime = \partial w / \partial z \ _{|z=0}$. 
The present bounds on $w_0^\prime$ \cite{Riess:2004nr}
allow a region of large $|w_0^\prime|$ for which the
validity of a Taylor expansion is doubtful even for $z=1$.

In addition to the dark
energy parameters $\od^0,\ w_0$ and $\ode$, we consider the matter
and baryon densities today $\omh$ and $\obh$, the Hubble parameter
$h$, optical depth to the last scattering surface $\tau$ and spectral
index $n$.

\begin{figure}
\vspace{1.2ex}
\includegraphics[height=0.45\textwidth,angle=-90]{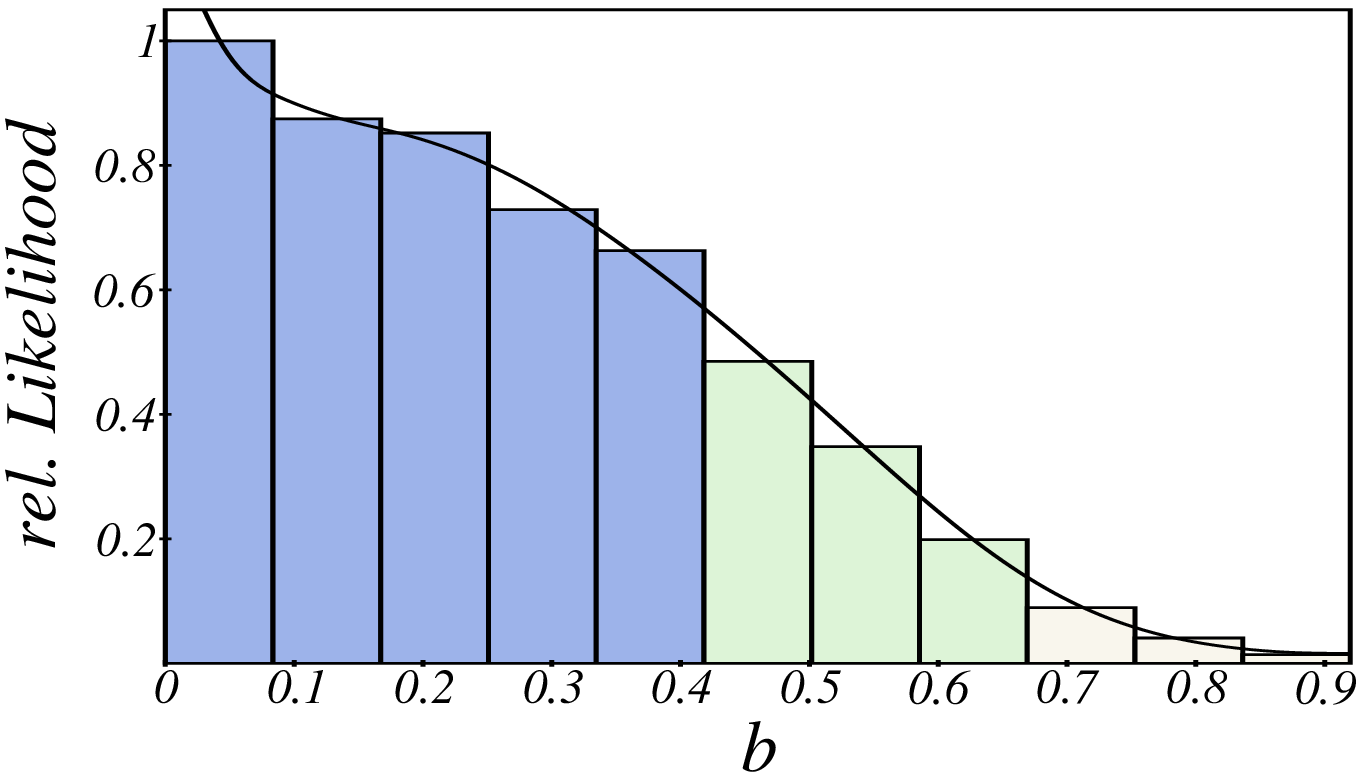}
\includegraphics[height=0.45\textwidth,angle=-90]{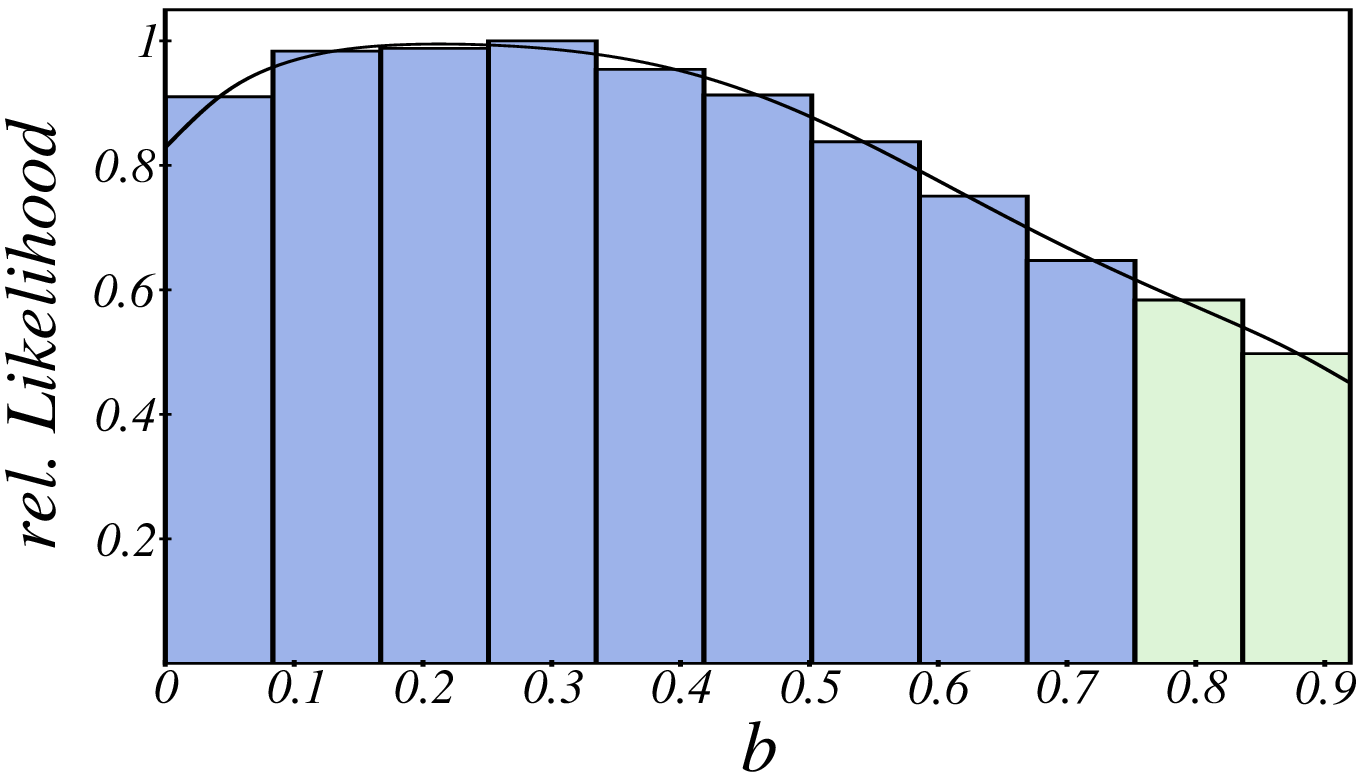}
\caption{Constraints on the parameter $b$ from WMAP+CBI+VSA+SDSS+HST data (upper panel)
and Sne Ia (lower panel).The dark (blue),
medium (green) and light (red) shaded regions correspond to $1,2$ and $3\sigma$
confidence. }
\label{fig::b}
\end{figure}

We compare the predictions of these models to the Sne Ia data of
\cite{Riess:2004nr} , as well as to the data from
WMAP \cite{Spergel:2003cb}, CBI \cite{Readhead:2004gy}, VSA
\cite{Rebolo:2004vp}, SDSS \cite{Tegmark:2003ud} and the Hubble
parameter constraint of the Hubble Space Telescope \cite{Freedman:2000cf}
combined. For this, we
employ the {\sc AnalyzeThis!} \cite{Doran:2003ua}  Monte Carlo package 
of  {\sc cmbeasy} \cite{Doran:2003sy}.  We ran simulations both in terms of 
$\ode, \od^0$ and $w_0$ as well as in terms of $b, \od^0$ and $w_0$, i.e
we computed chains where the parameter $\ode$ is traded for the parameter $b$ and
vice versa. For a constraint on the parameter $b$ from WMAP and Sne Ia, see 
Figure \ref{fig::b}.

\begin{figure}
\includegraphics[width=0.33\textwidth,angle=-90]{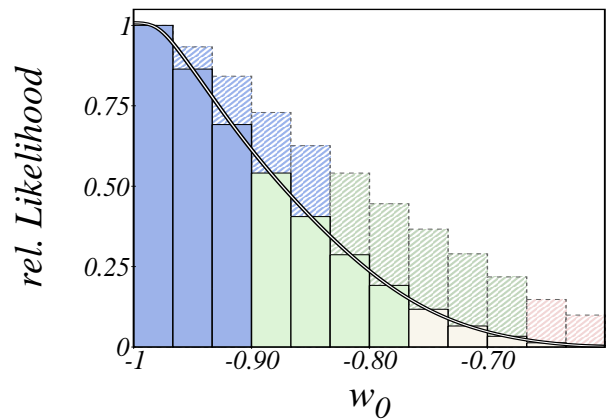}
\caption{Constraints on the equation of state today. The dark (blue),
medium (green) and light (red) shaded regions correspond to $1,2$ and $3\sigma$
confidence. The constraints from Sne Ia is displayed in the foreground (solid),
while the less tight constraint from WMAP+CBI+VSA+SDSS+HST is depicted in the
background.}
\label{fig::superimpw}
\end{figure}

\begin{figure}
\vspace{1.2ex}
\includegraphics[width=0.31\textwidth,angle=-90]{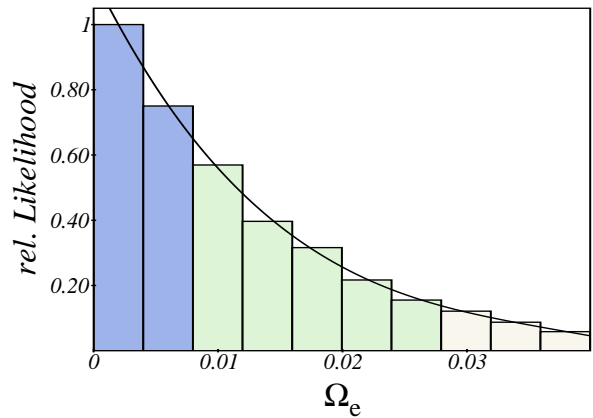}
\caption{Likelihood distribution of the early dark energy parameter $\ode$ inferred from
WMAP+CBI+VSA+SDSS+HST data. The dark (blue),
medium (green) and light (red) shaded regions correspond to $1,2$ and $3\sigma$
confidence. }
\label{fig::wmap_oe}
\end{figure}

As far as the model parameters $w_0$ and $\ode$ are concerned, the
main results are encapsulated in Figures  \ref{fig::oe_w}, \ref{fig::superimpw}
and \ref{fig::wmap_oe}. As is seen from Figure \ref{fig::oe_w}, the information provided by CMB plus
LSS and Sne Ia are almost orthogonal. While Sne Ia restrict $w_0$ considerably
better than CMB and LSS combined, the sensitivity to $\ode$ is worse.
From Sne Ia, we find $w_0 < -0.78$, while $\ode < 0.029$ from CMB, LSS  and HST.

\begin{figure}
\includegraphics[width=0.33\textwidth,angle=-90]{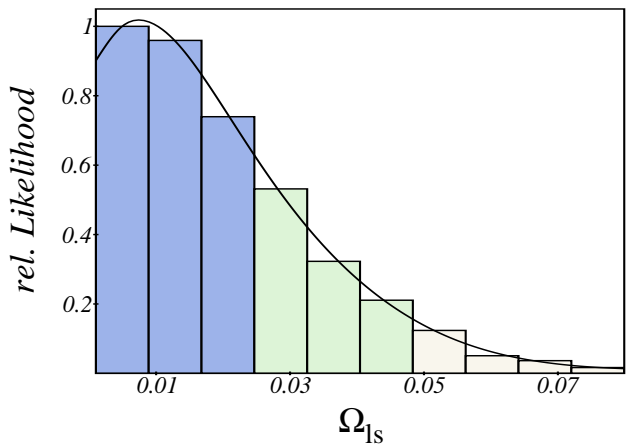}
\caption{The average dark energy contribution during structure formation $\ols$ as
defined in Equation \eqref{eqn::osf}. As $\ols$ is not a monte carlo parameter, but
a derived quantity, we plot the maximum likelihood of all models per bin. In the
model considered, we see that a model with overall Likelihood less than $\Delta \chi^2 =1$
apart from the best fit model can be found only for  $\ols \lesssim 3\%$. }
\label{fig::wmap_ols}
\end{figure}

\begin{figure}
\includegraphics[width=0.33\textwidth,angle=-90]{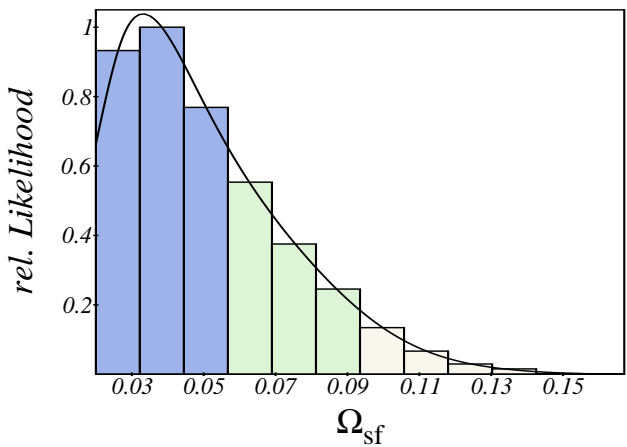}
\caption{The average dark energy contribution during structure formation $\osf$ as
defined in Equation \eqref{eqn::osf}. As $\osf$ is not a monte carlo parameter, but
a derived quantity, we plot the maximum likelihood of all models per bin. Please note that $\osf \sim 0.5\%$ 
by definition even for a cosmological constant Universe.}
\label{fig::wmap_osf}
\end{figure}

The average dark energy fractions at last scattering and during structure formation,
 $\ols$ and
$\osf$ can substantially deviate from $\ode$ (as seen in 
Figures \ref{fig::wmap_ols} and \ref{fig::wmap_osf}). As the underlying
probability density is unknown, we cannot quote constraints on 
$\ols$ and $\osf$ from counting the number of models in the chain
per bin. Instead, we choose to draw the distribution of the 
maximum likelihood model per bin. Figures  \ref{fig::wmap_ols} and  
\ref{fig::wmap_osf} show this distribution.
The somewhat tight $2-\sigma$ constraint on $\ols$ is a result of 
our parameterization: the amount of dark energy during structure
formation $\osf$ always exceeds $\ols$. As the ISW effect in the CMB
anisotropies as well as
effects on the cold dark matter power spectrum restrict $\osf \lessapprox 10\%$,
this leads to  $\ols \lessapprox 5\%$ at $95\%$ confidence level.

\begin{figure}
\includegraphics[width=0.31\textwidth,angle=-90]{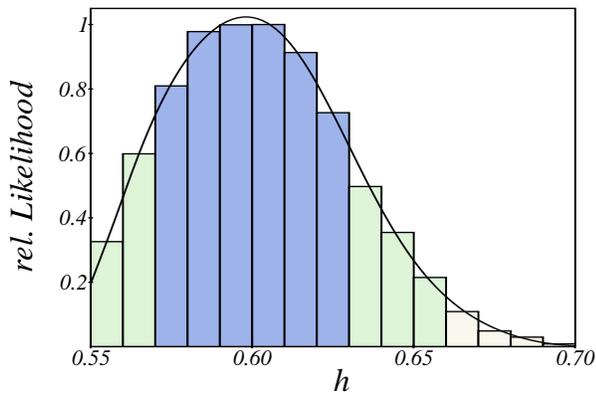}
\caption{Likelihood distribution for the hubble parameter $h$ yielding $h=0.6\pm0.03$. As explained
in the text, slightly lower values of $h$ than in standard $\Lambda$CDM models are preferred.
The data used was WMAP+CBI+VSA+SDSS+HST.}
\label{fig::hubble}
\end{figure}

Models of early dark energy typically lead to 
slightly lower values of $h$ than $\Lambda$-CDM. The
reason is the CMB:
dark energy influences the CMB mostly due to projection effects 
and additional ISW contributions. To match observations, any model must at least
provide an acoustic scale $l_A$ that comes close to the one measured
by WMAP. With $\od^0$ fixed, the analytic expression for $l_A$ of \cite{Doran:2000jt} 
yields 
\ee\label{eqn::integ}
l_A \propto \int_0^1 \textrm{d}a \left( a +
  \frac{\od^0}{1-\od^0} \, a ^{(1 - 3 \bar w)} +
  \frac{\Omega_{r}^0(1-a)}{1-\od^0} \right)^{-1/2}\hspace{-2em},
\eee
where $\Omega_r$ is the present energy fraction in radiation and $\bar w$ is the weighted average
\ee
\bar w =  \int_0^{\tau_0} \od(\tau) w(\tau) \textrm{d} \tau 
\times \left(  \int_0^{\tau_0} \od(\tau) \textrm{d} \tau
\right)^{-1}.
\eee
The integral \eqref{eqn::integ} increases when $\bar w \to -1$, i.e. for all 
other parameters fixed, the acoustic scale $l_A$ becomes larger the more negative $\bar w$. 
Conversely, as our models have $\bar w > -1$ by construction, we see
that for all other parameters fixed, our models have a smaller acoustic scale compared to
$\Lambda$-CDM. To counterbalance this, a somewhat smaller Hubble parameter $h$ is 
preferred (see Figure \ref{fig::hubble}), because  $l_A$ depends on $h$ 
to good approximation as \cite{Doran:2000jt} 
\ee
l_A \propto 1 + h^{-1} \sqrt{\frac{\Omega_{rel.}^0 h^2}{a_{ls}(1 - \od^0)}} \approx 1 + 0.4 h^{-1},
\eee
where we have used the estimates $\od^0 \approx 0.7$, $a_{ls} \approx 1100^{-1}$ and 
$\Omega_{rel.}^0 h^2 \approx 4.4\times 10^{-5}$.
Likewise, a sizeable early dark energy $\ols$ can increase the acoustic scale according
to    $l_A \propto  1/\sqrt{1-\ols} $  \cite{Doran:2000jt}.
Both $\ols$ and the somewhat smaller Hubble parameter counterbalance the effect of $\bar w$
in our models.

\begin{figure} 
\includegraphics[width=0.33\textwidth,angle=-90]{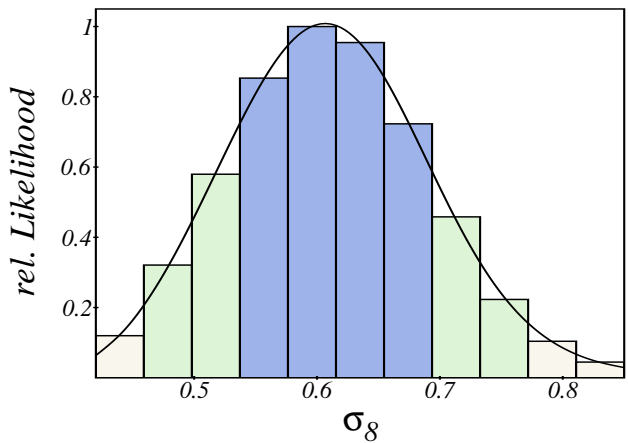}
\caption{The linear rms power of matter fluctuations on $8h^{-1} \mpc $ scales, $\sigma_8$. As $\sigma_8$
is not a monte carlo parameter, we caution the reader that the underlying prior
is not flat. With  $\sigma_8 = 0.61 \pm 0.08$ the power 
is rather low compared to a standard $\Lambda$ CDM model -- a direct result
of the suppression of growth due to the presence of dark energy during structure formation.}
\label{fig::wmap_sig8}
\end{figure}

Just as $\ols$ expresses the main effect of early dark energy on the CMB,
the suitable average
\ee\label{eqn::osf}
\osf = [\ln a_{tr.} - \ln a_{eq.}]^{-1} \int_{\ln a_{eq.}}^{\ln a_{tr.}} \od(a)\  \rd\ln a
\eee
with $a_{tr.} = 1/3$ encapsulates main effects of early dark energy on structure formation \cite{Doran:2001rw}.
By definition, one has $\osf(\Lambda) \sim 0.5\%$  (i.e. non-vanishing) for a cosmological constant model 
and our choice of $a_{tr.}$.
A sizeable $\osf$ leads to a decrease in linear structure compared to
$\Lambda$-CDM according to \cite{Doran:2001rw} 
\ee
\frac{\sigma_8(D.E.)}{\sigma_8(\Lambda)} \propto a_{eq.}^{3 \osf/5}
\eee 
Using $z_{eq.} = 3500$, we obtain for 
$3 \osf \ln z_{eq.} / 5 \ll 1$, the quick estimate
\ee
\frac{\sigma_8(D.E.)}{\sigma_8(\Lambda)} \propto (1 - 6\, [\osf - \osf(\Lambda)]),
\eee  
As we leave a free bias for our SDSS analysis, and have no prior
on $\sigma_8$, one may worry about unphysically low values  of
$\sigma_8$ for our models. It does turn out, however, that this is not
necessarily the case. As seen in Figure \ref{fig::wmap_sig8}, our predictions for
$\sigma_8$ are compatible with observations, given the
lower values of $\sigma_8$ needed to explain non-linear structure
in early dark energy scenarios compared to $\Lambda$CDM \cite{Bartelmann:2005fc}.

Cosmological probes have reached a level of accuracy that leads to
stringent constraints on any model of our Universe. In the case of 
the parameterization of dark energy in terms of $\od^0$, $\ode$ and $w_0$
we used, the Supernovae Ia and CMB+LSS experiments provide almost
orthogonal information. From the background evolution at $z \lesssim 2$ that
are probed by current Sne Ia observations, we find $w_0 \lesssim -0.78$ at 
$95$ \% confidence level, in good
agreement with prior investigations  (e.g. \cite{Hannestad:2002ur,Jassal:2004ej}).
 Likewise, structure formation and the cosmic
microwave background yield strong bounds on the abundance of 
dark energy at very high redshift. In our model, $\ode < 0.029$ at $95\%$ confidence level.
The value of $\ode$ is a factor of $2-3$ lower than the  abundance
of dark energy allowed during recombination $\ols$ and the abundance
during structure formation $\osf$. This is both a feature and a shortcoming
of our simple parameterization. Curing it would require more parameters which
at the present time are not necessary to explain current observations.
As early dark energy models such as the one investigated in this paper
make clear predictions for the CMB and in particular for structure formation,
it will become possible to falsify or confirm such a scenario within the
not so distant future. In this regard, the model investigated here holds
more opportunities for detection than models of dark energy with negligible
$\od$ at higher redshifts.

\begin{thebibliography}{}



\bibitem{Riess:2004nr}
A.~G.~Riess {\it et al.}  [Supernova Search Team Collaboration],
Astrophys.\ J.\  {\bf 607}, 665 (2004)
[arXiv:astro-ph/0402512].

\bibitem{Spergel:2003cb}
D.~N.~Spergel {\it et al.}  [WMAP Collaboration],
Astrophys.\ J.\ Suppl.\  {\bf 148}, 175 (2003)
[arXiv:astro-ph/0302209].

\bibitem{Readhead:2004gy}
A.~C.~S.~Readhead {\it et al.},
Astrophys.\ J.\  {\bf 609} (19??) 498
[arXiv:astro-ph/0402359].

\bibitem{Goldstein:2002gf}
J.~H.~Goldstein {\it et al.},
Astrophys.\ J.\  {\bf 599}, 773 (2003)
[arXiv:astro-ph/0212517].

\bibitem{Rebolo:2004vp}
R.~Rebolo {\it et al.},
arXiv:astro-ph/0402466.


\bibitem{Tegmark:2003ud}
M.~Tegmark {\it et al.}  [SDSS Collaboration],
Phys.\ Rev.\ D {\bf 69} (2004) 103501
[arXiv:astro-ph/0310723].


\bibitem{Hawkins:2002sg}
E.~Hawkins {\it et al.},
Mon.\ Not.\ Roy.\ Astron.\ Soc.\  {\bf 346} (2003) 78
[arXiv:astro-ph/0212375].

\bibitem{Freedman:2000cf}
  W.~L.~Freedman {\it et al.},
  Astrophys.\ J.\  {\bf 553} (2001) 47
  [arXiv:astro-ph/0012376].

\bibitem{Wetterich:fm}
C.~Wetterich,
Nucl.\ Phys.\ B {\bf{302}}, 668  (1988)



\bibitem{Ratra:1987rm}
B.~Ratra and P.~J.~Peebles,
Phys.\ Rev.\ D {\bf{37}}, 3406  (1988)

\bibitem{Caldwell:1997ii}
R.~R.~Caldwell,~R.~Dave and P.~J.~Steinhardt,
Phys.\ Rev.\ Lett.\  {\bf{80}}, 1582 (1998)

\bibitem{Caldwell:1999ew}
R.~R.~Caldwell,
Phys.\ Lett.\ B {\bf 545}, 23 (2002)
[arXiv:astro-ph/9908168].

\bibitem{Corasaniti:2002vg}
P.~S.~Corasaniti and E.~J.~Copeland,
Phys.\ Rev.\ D {\bf 67}, 063521 (2003)
[arXiv:astro-ph/0205544].

\bibitem{Linder:2002et}
E.~V.~Linder,
Phys.\ Rev.\ Lett.\  {\bf 90}, 091301 (2003).

\bibitem{Upadhye:2004hh}
  A.~Upadhye, M.~Ishak and P.~J.~Steinhardt,
  arXiv:astro-ph/0411803.

\bibitem{Wang:2004py}
  Y.~Wang and M.~Tegmark,
  Phys.\ Rev.\ Lett.\  {\bf 92}, 241302 (2004)
  [arXiv:astro-ph/0403292].






\bibitem{Doran:2000jt}
M.~Doran, M.~J.~Lilley, J.~Schwindt and C.~Wetterich,
Astrophys.\ J.\  {\bf 559}, 501 (2001)
[arXiv:astro-ph/0012139].


\bibitem{Doran:2001rw}
M.~Doran, J.~M.~Schwindt and C.~Wetterich,
Phys.\ Rev.\ D {\bf 64}, 123520 (2001)
[arXiv:astro-ph/0107525].


\bibitem{Caldwell:2003vp}
R.~R.~Caldwell, M.~Doran, C.~M.~Mueller, G.~Schaefer and C.~Wetterich,
Astrophys.\ J.\  {\bf 591} (2003) L75
[arXiv:astro-ph/0302505].



\bibitem{Wetterich:2004pv}
C.~Wetterich,
Phys.\ Lett.\ B {\bf 594}, 17 (2004)
[arXiv:astro-ph/0403289].



\bibitem{Doran:2003ua}
M.~Doran and C.~M.~Mueller,
J. Cosmol. Astropart. Phys. JCAP09(2004)003  [arXiv:astro-ph/0311311].


\bibitem{Doran:2003sy}
M.~Doran,
arXiv:astro-ph/0302138.



\bibitem{Bartelmann:2005fc}
  M.~Bartelmann, M.~Doran and C.~Wetterich,
  arXiv:astro-ph/0507257.
\bibitem{leibundgut}
B.~Leibundgut, private communication.




\bibitem{Jassal:2004ej}
  H.~K.~Jassal, J.~S.~Bagla and T.~Padmanabhan,
  Mon.\ Not.\ Roy.\ Astron.\ Soc.\  {\bf 356} (2005) L11
  [arXiv:astro-ph/0404378].

\bibitem{Hannestad:2002ur}
  S.~Hannestad and E.~Mortsell,
  Phys.\ Rev.\ D {\bf 66} (2002) 063508
  [arXiv:astro-ph/0205096].



\end{thebibliography}
\end{document}